\documentclass[twocolumn,showpacs,showkeys,preprintnumbers,amsmath,amssymb]{revtex4}

\usepackage{graphicx}
\usepackage{dcolumn}
\usepackage{bm}

\usepackage{color}

\begin{document}

\title{Periodic three-body orbits in the Coulomb potential} 

\author{Marija \v{S}indik$^{1}$}

\author{Ayumu Sugita$^{2}$}

\author{Milovan \v{S}uvakov$^{3}$}

\author{V. Dmitra\v{s}inovi\'{c}$^{3}$\footnote{to whom correspondence should be addressed, 
e-mail: dmitrasin@ipb.ac.rs}}

\affiliation{$^{1}$ Faculty of Physics, University of Belgrade,
Studentski Trg 12, 11000 Belgrade, Serbia} 
\affiliation{$^{2}$ Department of Applied Physics, Osaka City University, 
3-3-138 Sugimoto, Sumiyoshi-ku, Osaka 558-8585, Japan} 
\affiliation{$^{3}$ Institute of Physics Belgrade, University of Belgrade, Pregrevica 118, 
Zemun, 
11080 Beograd, Serbia}
\date{\today}







\begin{abstract}
We numerically discovered around 100 distinct non-relativistic collisionless periodic three-body orbits in the Coulomb potential 
{\it in vacuo}, with vanishing angular momentum, for equal-mass ions with equal absolute values of charges. These 
orbits are classified according to their symmetry and topology, and a linear relation is established between the 
periods, at equal energy, and the topologies of orbits. Coulombic three-body orbits can be formed in ion traps, such as 
the Paul, or the Penning one, where one can test the period vs. topology prediction.
\end{abstract}

\maketitle   

The Newtonian three-body problem is one of the outstanding classical open questions in science.
After more than 300 years of observation, only two topologically distinct types of 
periodic 
three-body systems, or orbits, have been observed in the skies \cite{Tokovinin:2007}: 
1) the so-called hierarchical systems, such as the Sun-Earth-Moon one, to which type 
belong more than 99\%  of all observed three-body systems; 2) Lagrangian three-body systems, 
such as Jupiter's Trojan satellites, to which the remaining $\leq 1\%$ belong. 

There has been some significant theoretical progress on the subject over the past few years: 
several hundred new, topologically distinct families of periodic solutions have been found 
by way of numerical simulations \cite{Moore1993,Chenciner:2000,Martynova2009,Suvakov:2013,Suvakov:2013b,Suvakov:2014,Iasko2014,Dmitrasinovic:2015,Dmitrasinovic:2017,site,Jankovic:2015,Rose:2016,Shibayama:2015,Li:2017,Liao:2017}, 
and unexpected regularities have been observed among 
them \cite{Dmitrasinovic:2015,Dmitrasinovic:2017,Liao:2017,Li:2017} relating the periods, 
topologies and linear stability of orbits. 

Of course, one would like to observe at least some of the new orbits and test their properties in 
an experiment, but such a test would be impeded by a number of 
obstacles: (1) only stable orbits have a chance of actually existing for a sufficiently long time 
to be observed; (2) stability depends on the ratio(s) of masses, and on the value of angular momentum, 
neither of which can be controlled in astronomical settings;
(3) even if an orbit is stable in a wide range of mass ratios and angular momenta, there is no 
guarantee that such a system will have been formed sufficiently frequently and sufficiently close  
to Earth, that it may be observed by our present-day instruments.
\begin{figure}[tbp]
\centerline{\includegraphics[width=0.95\columnwidth,,keepaspectratio]{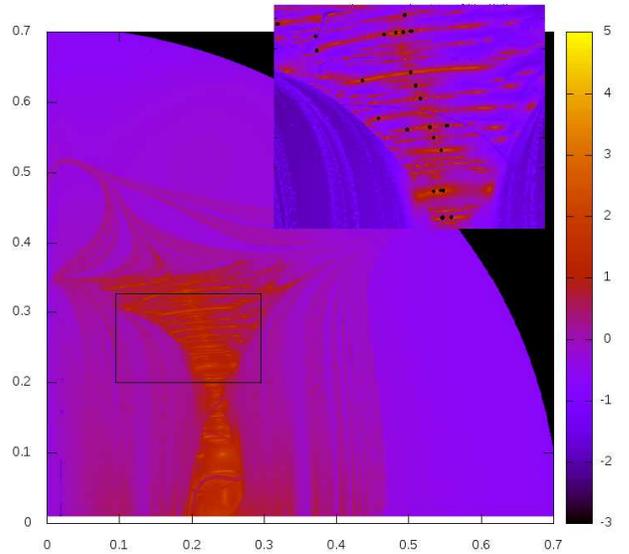}}
\caption{Negative logarithm of return proximity function as a function of the initial velocity
components $v_x$ and $v_y$ on the $x$ and $y$ axes, respectively. Bright areas correspond to 
high values of the negative logarithm. Inset: zoom-in of the ``boxed-in'' region.} 
\label{fig1}
\end{figure}

All of the above prompted us to look for alternative three-body systems that
share (at least) some of the same properties with Newtonian three-body systems.
The Coulombic potential shares one basic similarity with the Newtonian gravity -- its 
characteristic $1/r$ (homogeneous) spatial dependence -- as well as several important differences: 
(1) the (much) larger coupling constant; (2) both attractive and repulsive nature; 
(3) naturally identical
(quantized) electric charge(s); (4) ions with opposite charges may have masses equal to one part in a few 
thousand; (5) ions have a finite probability of elastic scattering in
head-on collisions; and (6) Coulombic bound states can be formed 
in table-top ion-trap experiments \cite{Major:2005}. 
For these reasons we turn to the study of periodic three-body orbits bound by Coulombic potential.
The application of only the Coulomb interaction amounts to a non-relativistic
approximation, which is good only in the low-velocity limit \cite{LandauII}.

In this Rapid Communication we present the results of a search that led to around 100 distinct collisionless 
orbits, only four of which are stable, and around 80 isosceles quasi-colliding (free-fall, or 
``brake'') ones. We use the collisionless orbits to display a new regularity, akin to Kepler's 
third law, in the form of a linear dependence 
\begin{equation}
T|E|^{3/2} \sim N,
\label{e:eq1}
\end{equation}
between the scale-invariant period $T|E|^{3/2}$, where $T$ is the period, 
and $E$ is the energy of an orbit, on one hand, and the orbit's topological complexity $N$,
expressed as the number of collinear configurations (``syzygies'') encountered during one cycle, 
see the text below, on the other. This prediction 
ought to be tested in ion trap experiments.

We used the same search method as in the Newtonian gravity three-body problem \cite{Suvakov:2013}.
There are 12 independent variables that define the initial state of this system, for 
each body there are the $x$ and $y$ coordinates of the body, and the $v_x$ and $v_y$ 
components of their velocity. Adopting the center-of-mass reference frame reduces 
this number (12) to eight. Fixing the value of angular momentum ($L=0$) reduces this further to six. 
Using the scaling rules \cite{Landau} for the solutions and the fact that periodic solution must pass through at least 
one syzygy (collinear configuration) during one period, yields a four-dimensional search space for 
all zero-angular-momentum periodic solutions. 
We search for solutions in the two-dimensional subspace of orbits that pass through the  
Euler configuration, defined as the symmetric collinear configuration wherein the positively 
charged particle with velocity $(-2v_x,-2v_y)$ passes through the origin $(0,0)$, i.e., exactly 
between the two negatively charged particles, which, in turn, pass through the points $(-1,0)$ and $(1,0)$,
both with velocity equal to $(v_x,v_y)$.

In order to search for periodic solutions numerically, we have discretized the search window in this 
two-dimensional subspace and calculated the return proximity function (RPF) $d_{T_0}(v_x, v_y)$ that 
measures how close to the initial condition the trajectory returns (see \cite{suppm}) up to some pre-defined 
upper limit on the integration time $T_0$, at each grid point. The negative logarithm of the computed RPF 
is shown in Fig. \ref{fig1}. 
Local minima of RPF are used as candidates for periodic solutions. After applying the gradient 
descent algorithm starting at each candidate point, we have declared as periodic the solutions 
with RPF of ${\cal O}(10^{-8})$ and smaller.

We have followed the example set by three-body orbits in Newtonian gravity \cite{Suvakov:2013,Dmitrasinovic:2017}, 
and classified the newly found Coulombic orbits according to their topologies, studied their 
stability and organized them into sequences some of which, though fewer in number, appear very 
similar to the Newtonian ones. 
Each orbit has a well defined topology which can be algebraized in at least two different ways, see \cite{suppm}
and Refs. \cite{Montgomery:1998,Tanikawa:2008}. Here we use Montgomery's method \cite{Montgomery:1998} 
wherein each solution is associated with (the conjugacy class, see \cite{Suvakov:2013}, of) an element 
of the two-generator (${\tt a}$, ${\tt b}$, ${\tt A=a^{-1}}$, ${\tt B=b^{-1}}$) free group $F_2({\tt a, b})$. 

There are important distinctions among the 100-odd orbits: 
1) the orbits can be separated into two classes, using their symmetry: class (A) consists of 
orbits that are symmetrical under two perpendicular reflections, and class (B) of orbits with a 
point reflection symmetry; 
2) each of the classes can be further separated into sequences, defined by their free-group elements,
as follows. For both class A and class B, sequence (I): 
$w_{n,k}^{\rm {(I)}} = {\tt [(AB)^n (ab)^n]^{k}}$ with integers $n,k=1,2,\ldots$;
and for class A only, sequence (II): $w_{m,n,k}^{\rm {(A.II)}} = {\tt [(AB)^m (ab)^n]^{k} A [(BA)^m (ba)^n]^{k} B}$, 
with $m,n,k = 1,2,3, \cdots$; and sequence (III): 
$w_{n}^{\rm {(III)}} ={\tt [(ab)^2  ABA (ba)^2 BAB]^{n}}$, with $n = 1,2,3, \cdots$.

Note that the 100-odd collisionless Coulombic orbits are substantially fewer than roughly 200 collisionless 
Newtonian orbits with similar search parameters, and that there are only four linearly 
stable solutions in contrast to more than 20 in the Newtonian case.

All of this is a consequence of just one sign change in the potential: one pair of charged particles 
must experience repulsion, contrary to Newtonian gravity, where all pairs are attractive. Therefore, 
no choreographic solution, i.e. permutationally symmetric solution with all three particles following 
the same trajectory, such as the famous ``figure-8'' orbit, may exist in the Coulombic case. Moreover, 
at least one orbit, similar to Orlov's \cite{Martynova2009} colliding ``$S$-orbit'' (in Newtonian gravity) 
still exists in the Coulombic case, but it is not stable any more, and consequently does not 
produce an infinite sequence of periodic orbits, see \cite{Dmitrasinovic:2017}.

\begin{figure*}[tbp]
\vskip -30pt
\centerline{\hskip 70pt \includegraphics[width=7in,,keepaspectratio]{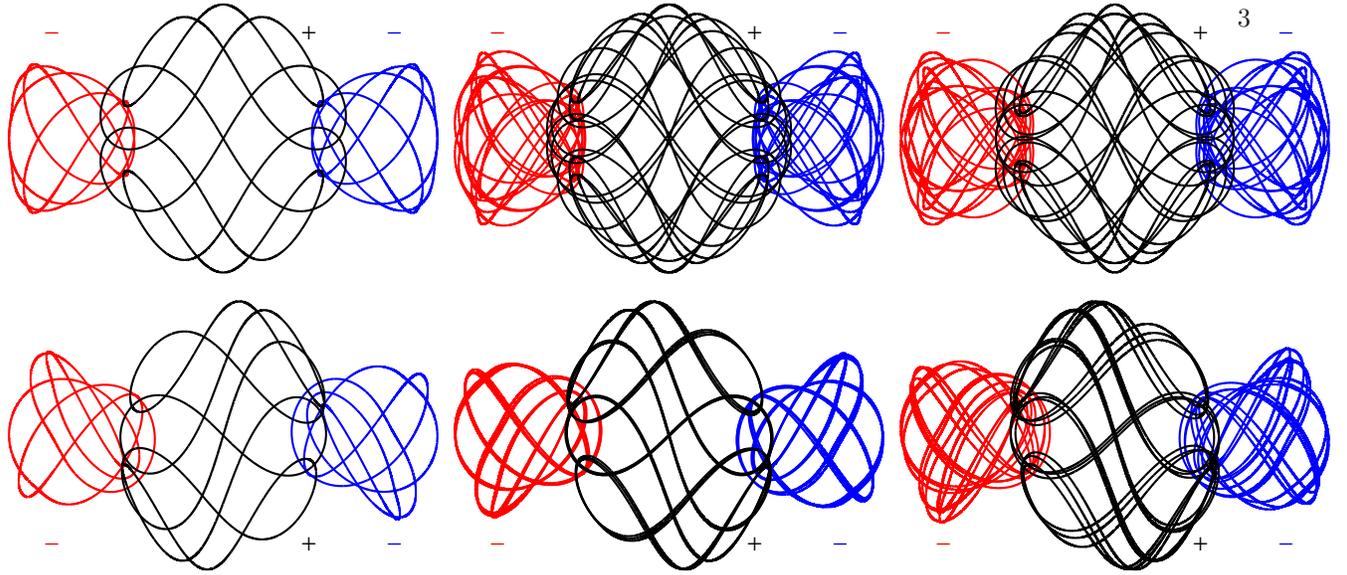}}
\vskip 30pt
\caption{Trajectories of orbits A.4, with topology ${\tt (AB)^2 (ab)^2}$; A.12.a and A.12.b, both with topology
${\tt [(AB)^2 (ab)^2]^{3}}$, in class A (upper row); and orbits B.4, with topology ${\tt (AB)^2 (ab)^2}$,
and B.12.a and B.12.b, both with topology ${\tt [(AB)^2 (ab)^2]^{3}}$, in class B (lower row), respectively. 
Note the independent symmetries of the class A (upper row) trajectories with respect to the 
reflections about the horizontal and the vertical axis, whereas the class B (lower row) trajectories
have only this symmetry under combined reflections. Black lines correspond to the positively charged 
particle while red and blue lines correspond to the negatively charged ones.}
\label{fBIG}
\end{figure*}

\begin{table*}[tbh]
\centering
\caption{Initial conditions of six orbits, depicted in Fig. \ref{fBIG}, that belong to the sequence 
described by the free-group elements ${\tt [(AB)^2 (ab)^2]^{k}}$, with $k = 1,2,3, \cdots$, and 
four linearly stable orbits, Table \ref{tab:stabilityI}. The columns correspond to: solution label, 
name of the sequence that the solution belongs to, initial velocities 
$[\dot{x_1}(0)$, and $\dot{y_1}(0)]$, period, negative energy, scaled period, free group element, number 
of letters in free group element (equal to number of asymmetric syzygies), and the total number of syzygies 
over a period. For initial conditions of all other found solutions, see \cite{suppm}.}
\begin{tabular}{l|c|c|c|c|c|c|c|c|c}
\hline \hline
Label & Seq. & $\dot{x_1}(0)$    & $\dot{y_1}(0)$   & $T$  & $-E$  & $T|E|^{3/2}$  & Free group element & $N_w$ & $N_e$ \\ 
\hline
A.4 & I & 0.191764 & 0.330958 & 13.4332 & 1.06108 & 14.6826 & ${\tt (AB)^2 (ab)^2}$  & 8 & 14 \\
A.12.a & I & 0.147917 & 0.323693 & 37.1599 & 1.12003 & 44.0473 & ${\tt [(AB)^2 (ab)^2]^3}$ & 24 & 42 \\
A.12.b & I & 0.246251 & 0.335527 & 45.3784 & 0.980345 & 44.0472 & ${\tt [(AB)^2 (ab)^2]^3}$ & 24 & 42\\
\hline
B.4 & I & 0.111427 & 0.305087 & 11.3981 & 1.18352 & 14.6755 & ${\tt (AB)^2 (ab)^2}$ & 8 & 14 \\
B.12.a & I & 0.327539 & 0.337033 & 57.4554 & 0.83738 & 44.0266 & ${\tt [(AB)^2 (ab)^2]^3}$ & 24 & 42\\
B.12.b & I & 0.345214 & 0.344247 & 63.0644 & 0.786962 & 44.0266 & ${\tt [(AB)^2 (ab)^2]^3}$ & 24 & 42\\
\hline 
A.15.b & II & 0.108065 & 0.323579 & 44.7536 & 1.15086 & 55.2534 & ${\tt (ab)^2 ABA (ba)^2 b ABA }$ &  \\
& & & & & & & $ \times {\tt (ba)^2 BAB (ab)^2a BAB}$  & 30  & 52\\
A.18 & III & 0.105224 & 0.336995 & 55.6513 & 1.12609 & 66.5019 & ${\tt (ab)^2 ABA (ba)^2 (BA)^2 bab}$ &  \\
& & & & & & & $ \times {\tt (AB)^2 aba (BA)^2 (ba)^2 BAB}$  & 36 & 62  \\
A.20.b & I & 0.126494 & 0.315968 & 59.3293 & 1.15249 & 73.4049  & ${\tt [(ab)^2 (AB)^2]^5}$ & 40 & 70 \\
A.24.a & II & 0.249577 & 0.291337 & 80.2223 & 1.0585 & 87.364 & ${\tt [(ab)^2 (AB)^2 A]^2 (ba)^2 b}$ & \\
&  & & &  & & & $ \times {\tt (AB)^2 [(ab)^2 a (BA)^2 B]^2}$ & 48 & 86 \\
\hline
\hline
\end{tabular}
\label{tab:Coulomb_A_B}
\end{table*}

The initial conditions of all 100-odd orbits and their corresponding topological and kinematical 
properties can be found in \cite{suppm}; in Fig. \ref{fBIG} and Table \ref{tab:Coulomb_A_B}
we have shown six representative solutions.

\begin{figure}[tbp]
\vskip -30pt
\centerline{\hskip 50pt \includegraphics[width=0.95\columnwidth,,keepaspectratio]{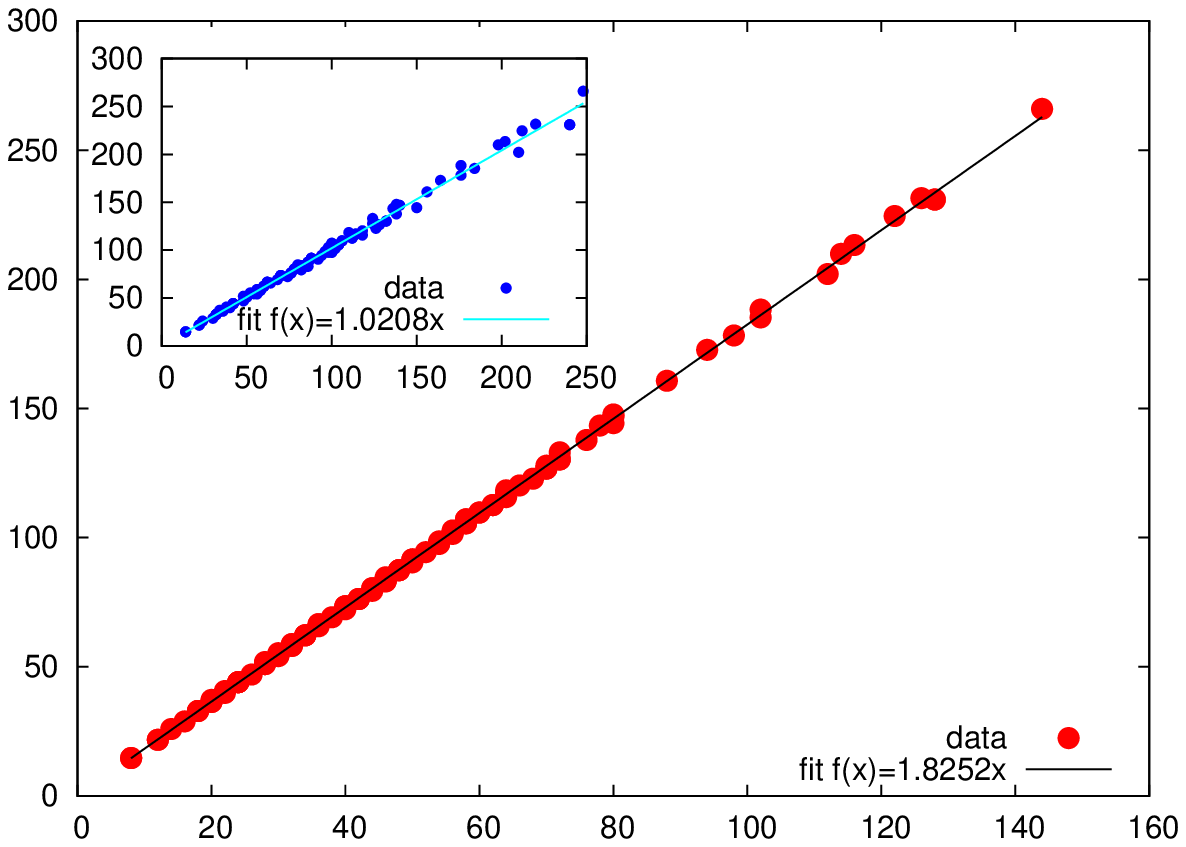}}
\vskip 20pt
\caption{Dependence of the scale-invariant period $T|E|^{3/2}$ on the number of asymmetric syzygies $N_w$ (collinear 
configurations with two particles of the same charge on one side) during an orbit. Inset: dependence of 
the scale-invariant period $T|E|^{3/2}$ on the number of all syzygies $N_e$.} 
\label{fig3}
\end{figure}

Next we show that Eq. (1), the (striking) property of orbits that was first observed in Newtonian three-body 
systems \cite{Dmitrasinovic:2015}, also features in the Coulombic three-body systems.
This relation between topological and kinematical properties of Newtonian three-body 
systems was first reported in \cite{Dmitrasinovic:2015} and later studied in more detail in Refs. 
\cite{Jankovic:2015,Liao:2017,Li:2017,Dmitrasinovic:2017}. 
Equstion (1) is a (simple) linear dependence of the scale-invariant period $T|E|^{3/2}$ on the topological complexity $N$. 
The topological complexity $N$ can be measured in at least two different ways: (1) we used the length $N_w$ of the 
free-group element (word) describing the orbit's topology, which, due to symmetry in our case, is equal to the number of 
asymmetric syzygies, i.e., collinear configurations wherein the two equal-charge particles are next to each other, 
over one period; (2) the number $N_e$ of {\it all} syzygies (collinear configurations)) was 
considered in Refs. \cite{Dmitrasinovic:2015,Dmitrasinovic:2017} as the measure of 
topological complexity $N$ of Newtonian orbits. 

In Fig. \ref{fig3} one can see that Eq. (1) holds for three-body orbits in the Coulomb potential: 
(1) with $N=N_w$, a linear fit yields a slope equal to 1.8252, with asymptotic standard error of 
0.08\% and an average relative deviation of points from fit values that equals 0.63\%; 
(2) with $N=N_e$ the number of {\it all} syzygies (collinear configurations), 
in the Coulomb case, the situation is slightly different (see inset in Fig. \ref{fig3}): the slope 
of this fit is $1.0208$, the asymptotic standard error is 0.36\%, but with a significantly larger 
(2.6\%) average relative deviation of points from fit values.

As mentioned earlier, only four solutions are linearly stable. We solved the equations for an 
infinitesimal deviation from the exact periodic solution along each periodic orbit 
to find the eigenvalues of the monodromy matrix, see \cite{suppm}.
Due to the symmetry of the equations of motion, these eigenvalues appear as two ($i=1,2$) quadruples ($\lambda_i$, $\lambda_i^{*}$, 
$1/\lambda_i$, $1/\lambda_i^{*}$). For only four orbits listed in Table \ref{tab:stabilityI} both eigenvalues $\lambda_i$ 
have moduli equal to unity $|\lambda_i|=1$, within their respective margins of error, which means that the
corresponding three-body orbit is linearly stable.

\begin{table}[tbh]
\centering
\caption{Stability coefficients $\lambda_j,\nu_j$, with $j=1,2$,  
of linearly stable (double elliptic) orbits, where $\lambda_j = \exp(2 \pi i \nu_j)$.} 
\begin{tabular}{l|c|c|c|c|c}
\hline \hline
name & ${\rm Re}(\lambda_j)$  & ${\rm Im}(\lambda_j)$  & $|\lambda_j|^2$ & 
$\nu_j$ & $N_w$ \\ 
\hline
A.15.b &  0.510145  & 0.860102 & 1.000023  & 0.164797 & 30 \\
 & -0.11507  & 0.993357 & 0.999999  & 0.268355 &  \\
\hline
A.18  & -0.002025 & 0.999961 & 0.999926  & 0.250322 & 36 \\
  & -0.820340 & 0.571882 & 1.000007  & 0.403107 & \\
\hline
A.20.b & 0.009875 & 0.998966 & 0.998031  & 0.248427 & 40 \\
 & 0.94189  & 0.339728 & 1.002572  & 0.055094 & 
 \\
\hline
A.24.a &  -0.988601 & 0.174067 & 1.007631  & 0.472261 & 48 \\
&  0.993975  & 0.116863 & 1.001643  & 0.018627 & 
\\
\hline
\end{tabular}
\label{tab:stabilityI}
\end{table}

Thus we have shown that some of the phenomena first observed in Newtonian 
three-body orbits, such as the linear dependence of the scale-invariant period on the topology 
of the orbit, 
and the emergence of sequences \footnote{from a linearly stable orbit, in agreement with the 
Birkhoff-Lewis 
theorem \cite{Birkhoff_Lewis:1933}} exist in Coulombic three-body orbits, and are not features of 
Newtonian gravity alone. The homogeneity \cite{Landau}  of the Coulombic, Newtonian, and the 
strong Jacobi-Poincar\'e potentials is common to all three known cases of manifestation of this
regularity \cite{Landau,Dmitrasinovic:2017,Dmitrasinovic:2017b}. This supports indirectly the 
explanation offered in Refs. \cite{Dmitrasinovic:2017,Dmitrasinovic:2017b}.
 
Our next concern ought to be the observation of some of these orbits in an experiment. 
The trajectories of a number (ranging between 1 and 32) of positively charged particles
moving in a Paul trap have been photographically recorded as early as 1959 \cite{Major:2005,Wuerker:1959}. 
The challenge to actually confine and photograph a few oppositely charged macroscopic particles
in an ion trap has remained unanswered to the present day, to our knowledge. It is well known that Paul and/or Penning
traps can lead to binding of pairs of identical ions, including periodic orbits as well as their chaotic motions 
\cite{Hoffnagle:1988,Totsuji:2002}, when the circumstances (such as the frequency and amplitudes 
of the applied electric and/or magnetic fields) are right. Such
periodic orbits are impossible in free space, however, as there
the identical ions experience only Coulomb repulsion 
\footnote{Curiously, the case of two opposite-charged, equal-mass 
ions has not been examined in detail, thus far, see e.g. \cite{Trypogeorgos:2013}, although it 
also falls into the first Mathieu stability zone.}
So, before one observes any periodic three-body orbits in an
ion trap, and declares them genuine Coulomb orbits, one must
know which periodic three-body orbits exist in free space -- information that we have provided 
here.
With the present work we have prepared the terrain for future numerical, and we hope also 
experimental studies of three-ion motions in traps \cite{Major:2005}. 

Naturally, the orbits that are (linearly) stable in free space are also expected to exist in a trap; 
that is not to say that the unstable orbits cannot be stabilized by appropriate trapping fields,
or that new kinds of periodic orbits cannot be formed in a trap. Moreover, ions 
have a non-zero elastic head-on collision cross-section, unlike the stars and/or planets, so one may even  
observe some ``colliding'' orbits \footnote{We put the word colliding into quotation marks here, 
because only zero-angular momentum orbits experience actual collisions, whereas ion traps generally 
impart angular momentum to ions.} 
in ion traps. This gives one an opportunity to observe hitherto 
experimentally unobserved orbits and to study some of their unprecedented properties. 

At any rate, trap-induced corrections will have to be calculated for each three-ion orbit 
in any trap where experiments are conducted, before an interpretation is given.  
With this Rapid Communication we hope to start a discussion of trap-induced corrections for periodic three-ion orbits: 
in order to calculate such corrections, one needs the (initial conditions of) free-space periodic orbits, 
of which we have provided around 100, that ought to suffice for a starting point.

There are no records, to our knowledge, of searches for periodic Coulombic three-body systems 
with equal masses and equal charges, which are the closest to the equal-mass Newtonian system 
that was studied in Refs. \cite{Martynova2009,Suvakov:2013,Suvakov:2013b,Suvakov:2014,Iasko2014,Dmitrasinovic:2015,site,Jankovic:2015,Rose:2016,Shibayama:2015,Li:2017,Liao:2017}.
As we wished to compare the closest analogons of the Coulombic and Newtonian three-body systems, 
we had to repeat a search for periodic collisionless orbits at the present mass and charge ratios.

To be sure, we are not the first ones who have studied Coulombic periodic three-body motion: 
the subject has a long history, see e.g. Refs. \cite{Langmuir:1921,Wannier:1953}, 
with a revival in the 1980s, since when a number of studies have been published:  
\cite{DAVIES:1983,Klar:1986a,Poirier:1989,Grujic:1988,Richter:1990,Richter:1993,Yamamoto:1993,Santander:1997,Stuchi:2000,Sano:2004,Rupp:2010}.
Numerical discovery of more than 8000 collinear colliding periodic orbits with He atom mass ratios was 
reported in Ref. \cite{Richter:1993}, and of somewhat fewer collisionless ones in Ref. \cite{Yamamoto:1993}. 
The initial conditions were not published, so one could not simply retrieve these previously 
discovered orbits and use them here.

With this Rapid Communication we also hope to induce practitioners to consider experimental 
searches, particularly in view of the fact that, at least in the case of past periodic-orbit 
discoveries, the theory did not precede experiment \cite{Major:2005}.

The work of V. D. and M. \v Suvakov was supported by the Serbian Ministry of Education, 
Science and Technological Development under Grants No. OI 171037 and No. III 41011.
M. \v Sindik conducted her work during the summer breaks of 2016 and 2017, when
she was supported by a Student scholarship from the Serbian Ministry of Education, 
Science and Technological Development.
All numerical work was done on the Zefram cluster, Laboratory for gaseous electronics,
Center for non-equilibrium processes, at the Institute of Physics, Belgrade.
We thank Marija Jankovi\' c, Ana Hudomal and Srdjan Marjanovi\' c for their help 
in the early stages of this work.


\begin{thebibliography}{20}


\bibitem{Tokovinin:2007}
A. Tokovinin, ``Dynamics of Multiple Stars: Observations'',
{\it Massive Stars in Interacting Binaries}, ASP Conference Series, Vol. 367, 
edited by N. St-Louis \& A.F.J. Moffat, Astronomical Society of the Pacific, San Francisco, (2007),
pp. 615-619.

\bibitem{Moore1993}
C. Moore, ``Braids in classical gravity'',
Phys. Rev. Lett. {\bf 70}, 3675 - 3679 (1993).

\bibitem{Chenciner:2000}
A. Chenciner and R. Montgomery, 
``A remarkable periodic solution of the three-body problem in the case of equal masses,''
Ann. Math. {\bf 152}, 881--901 (2000).

\bibitem{Martynova2009}
A. I. Martynova, V. V. Orlov, and A. V. Rubinov,
Astron. Rep. {\bf 53}, 710 (2009).

\bibitem{Suvakov:2013}
M.~\v Suvakov, and V.~Dmitra\v sinovi\' c, ``Three classes of Newtonian three-body planar periodic orbits'', 
Phys.\ Rev.\ Lett. {\bf 110}, 114301 (2013).

\bibitem{Suvakov:2013b}
M. ~\v Suvakov, ``Numerical search for periodic solutions in the vicinity of the
figure-eight orbit: Slaloming around singularities on the shape sphere,'' 
Celest. Mech. Dyn. Astron. {\bf 119}, 369-377 (2014). 

\bibitem{Suvakov:2014}
M.~\v Suvakov, and V.~Dmitra\v sinovi\' c, ``A guide to hunting periodic three-body orbits'',  
Am. J. Phys. {\bf 82}, 609-619 (2014).
  
\bibitem{Shibayama:2015}
Milovan ~\v Suvakov and Mitsuru Shibayama, 
``Three topologicaly nontrivial choreographic motions of three bodies'', 
Celest. Mech. Dyn. Astron. {\bf 124}, 155-162 (2016).

\bibitem{Dmitrasinovic:2015}
V. Dmitra\v{s}inovi\'{c} and M. \v{S}uvakov,
``Topological Dependence of Kepler's Third Law for Planar Periodic 
Three-Body Orbits with Vanishing Angular Momentum'',
Phys. Lett. {\bf A 379}, 1939-1945 (2015).

\bibitem{Iasko2014}
P. P. Iasko and V. V. Orlov,``Search for Periodic Orbits in the General Three-Body Problem'', 
Astron. Rep. {\bf 58}, 869-879. (2014).

\bibitem{Jankovic:2015}
M. R. Jankovi{\' c} and V. Dmitra\v sinovi{\' c}, 
``Angular momentum and topological dependence of Kepler's Third Law in 
the Broucke-Hadjidemetriou-H\' enon family of periodic three-body orbits'', 
Phys. Rev. Lett. {\bf 116}, 064301 (2016).

\bibitem{Rose:2016}
Danya Rose, Ph.D. thesis 
``Geometric phase and periodic orbits of the equal-mass, 
planar three-body problem with vanishing angular momentum'',
University of Sydney, 2016. Available at {\tt https://ses.library.usyd.edu.au/handle/2123/14416}
 
\bibitem{Dmitrasinovic:2017}
V. Dmitra\v{s}inovi\'{c}, Ana Hudomal, Mitsuru Shibayama, Ayumu Sugita,
``Linear Stability of Periodic Three-Body Orbits with Zero Angular Momentum and Topological Dependence of 
Kepler's Third Law: A Numerical Test'', J. Phys. A: Math. Theor. {\bf 51}, 315101 (2018),

\bibitem{site}
{\tt http://three-body.ipb.ac.rs/}
and 
{\tt http://three-body.ipb.ac.rs/sequences.php}

\bibitem{Liao:2017}
Xiaoming Li and Shijun Liao, 
``More than six hundred new families of Newtonian periodic planar collisionless three-body orbits'', 
Sci. China-Phys. Mech. Astro. Vol. 60 No. 12: 129511 (2017). 

\bibitem{Li:2017}
X. Li, Y. Jing, S. Liao, 
``The 1223 new periodic orbits of planar three-body problem with unequal mass and zero angular momentum'',
Publ. Astron. Soc. Japan, {\bf 70}, psy057, 1-7 (2018).

\bibitem{Major:2005}
Fouad G. Major, Viorica N. Gheorghe, G\"unter Werth, 
``Charged Particle Traps: Physics and Techniques of Charged Particle Field Confinement'', 
Springer Berlin Heidelberg New York (2005).

\bibitem{LandauII}
L.D. Landau and E. M. Lifshitz, {\it The Classical Theory of Fields}, 
(4th revised English ed.) Butterworth-Heinemann,  Oxford (1994).

\bibitem{Landau}
Sect. 10, ``Mechanical Similarity'' in L.D. Landau and E. M. Lifshitz, {\it Mechanics}, 
(3rd ed.) Butterworth-Heinemann,  Oxford (1976).

\bibitem{suppm}
See Supplemental Material at [URL will be inserted by publisher] for Periodic three-body orbits 
in the Coulomb potential, which includes, but is not limited to: (a) tables of initial conditions,
topologies of orbits; (b) figures of orbits' tajectories; (c) description of the search method;
(d) stability analysis.
 
\bibitem{Montgomery:1998}
R. Montgomery, ``The N-body problem, the braid group, and action-minimizing periodic solutions'',
Nonlinearity {\bf 11}, 363 - 376 (1998).




\bibitem{Tanikawa:2008}
K. Tanikawa and S. Mikkola, ``A trial symbolic dynamics of the planar three-body problem'',
Proc. ``Resonances, Stabilization, and Stable Chaos in Hierarchical Triple Systems'', 
ed. V. V. Orlov \& A. V. Rubinov, St Petersburg State University Press, St Petersburg, (2008), 
pp. 26. (arXiv:0802.2465) 

\bibitem{Birkhoff_Lewis:1933}
G. D. Birkhoff and  D. C. Lewis, ``On the periodic motions near a given periodic motion 
of a dynamical system'', Ann. Mat. Pura Appl. {\bf 12}, 117-133 (1933).

\bibitem{Dmitrasinovic:2017b}
V. Dmitra\v sinovi\' c, Luka V. Petrovi\' c and Milovan \v Suvakov,
``Periodic three-body orbits with vanishing angular momentum in the pairwise ``strong'' potential'',  
Jour. Phys. {\bf A 50}, 435102 (2017).

\bibitem{Wuerker:1959}
R. F. Wuerker, H. Shelton, and R. V. Langmuir, ``Electrodynamic containment of charged particles,'' 
J. Appl. Phys. 30, 342 (1959).

\bibitem{Hoffnagle:1988}
J. Hoffnagle, R. G. DeVoe, L. Reyna and R. G. Brewer, 
``Order-Chaos Transition of Two Trapped Ions'', Phys. Rev. Lett., {\bf 61}, 255 (1988).

\bibitem{Totsuji:2002}
Hiroo Totsuji, Tokunari Kishimoto, Chieko Totsuji, and Kenji Tsuruta
``Competition between Two Forms of Ordering in Finite Coulomb Clusters'',
Phys. Rev. Lett., {\bf 88}, 125002 (2002).

\bibitem{Trypogeorgos:2013}
Trypogeorgos, D., Foot, C.: 
``Cotrapping different species in ion traps using multiple radio frequencies''
Phys. Rev. {\bf A 94}, 023609 (2016); arXiv:1310.6294 (2013).

\bibitem{Langmuir:1921}
Irving Langmuir, ``The Structure of the Helium Atom'', Phys. Rev. {\bf 17}, 339 (1921).


\bibitem{Wannier:1953}
Gregory H. Wannier, ``The Threshold Law for Single Ionization of Atoms or Ions by Electrons'',
Phys. Rev. {\bf 90}, 817 (1953).

\bibitem{DAVIES:1983}
I. Davies, A. Truman, and D. Williams, 
``Classical Periodic Solutions of the Equal-Mass 2n-Body Problem,
2n-Ion Problem and the n-Electron Atom Problem'',
Phys. Lett. {\bf 99 A}, 15 (1983). 


\bibitem{Klar:1986a}
H. Klar, ``Equilibrium Atomic Structure: Rotating Atoms''
Phys. Rev. Lett. {\bf 57}, 66 (1986).



\bibitem{Poirier:1989}
M. Poirier, Phys. Rev. {\bf A 40}, 3498 (1989).

\bibitem{Grujic:1988}
P. Gruji\'c and N. Simonovi\'c, ``The classical helium atom -- An asynchronous-mode model'', 
J. Phys. {\bf B 24}, 5055 (1991).


\bibitem{Richter:1990}
K. Richter and D. Wintgen, ``Stable Planetary Atom Configurations'',
Phys. Rev. Lett. 65, 1965 (1990).



\bibitem{Richter:1993}
K. Richter, G. Tanner and D. Wintgen, ``Classical mechanics of two-electron atoms''
Phys. Rev. {\bf A 48}, 4182 (1993). 


\bibitem{Yamamoto:1993}
Tomoyuki Yamamoto and Kunihiko Kaneko, ``Helium Atom as a Classical Three-Body Problem''
Phys. Rev. Let. {\bf 70}, 1928 (1993). 



\bibitem{Santander:1997}
A. Santander, J. Mahecha, and F. P\'erez, ``Rigid-Rotator and Fixed-Shape Solutions to the
Coulomb Three-Body Problem'', Few-Body Systems {\bf 22}, 37-60 (1997).



\bibitem{Stuchi:2000}
T.J. Stuchi, A.C.B. Antunes, and M.A. Andreu, 
``Muonic Helium atom as a classical three-body problem'', 
Phys. Rev. {\bf E 62}, 7831-7841 (2000).

\bibitem{Sano:2004}
M.M. Sano, ``The classical Coulomb three-body problem in the collinear eZe configuration'', 
J. Phys. {\bf A 37}, 803-822 (2004).

\bibitem{Rupp:2010}
Florian Rupp, J\"urgen Scheurle, ``Genuine Equilibria of Three Body Coulomb Configurations'',
Few-Body Syst {\bf 48}, 1-10 (2010).

\end{thebibliography}
\end{document}